# Real-space determination of orbital states driving successive phase transitions in $FeV_2O_4$

Chihaya Koyama[1,*], Yusuke Nomura[2,3], Shunsuke Kitou[1,†], Taishun Manjo[4], Yuiga Nakamura[4], Takeshi Hara[5], Naoyuki Katayama[6], Yoichi Nii[7], Ryotaro Arita[8,9], Hiroshi Sawa[10], and Taka-hisa Arima[1,9]

[1] *Department of Advanced Materials Science, The University of Tokyo, Kashiwa 277-8561, Japan.*
[2]*Institute for Materials Research (IMR), Tohoku University, Sendai 980-8577, Japan.*
[3]*International Quantum Materials Center (Q²), Advanced Institute for Materials Research (WPI-AIMR), Tohoku University, Sendai 980-8577, Japan.*
[4]*Japan Synchrotron Radiation Research Institute (JASRI), SPring-8; Hyogo 679-5198, Japan.*
[5]*Department of Physics, Tohoku University, Sendai 980-8578, Japan.*
[6]*Department of Physics, Okayama University, Okayama 700-8530, Japan.*
[7]*Department of Applied Physics and Physico-Informatics, Keio University, Kanagawa 223-8522, Japan.*
[8]*Department of Physics, The University of Tokyo, Tokyo 153-8904, Japan.*
[9]*RIKEN Center for Emergent Matter Science (CEMS), Wako 351-0198, Japan.*
[10]*Nagoya Industrial Science Research Institute, Nagoya 460-0008, Japan.*

[*]koyama-chihaya@g.ecc.u-tokyo.ac.jp
[†]kitou@edu.k.u-tokyo.ac.jp

**Abstract**

Direct experimental access to orbital states in strongly correlated materials remains a major challenge, despite their central role in driving coupled structural and magnetic phase transitions. In systems where electronic correlations, electron–lattice coupling, and relativistic spin–orbit interactions compete on comparable energy scales, first-principles calculations often yield multiple metastable solutions, hindering the unambiguous identification of the ground state. Here, we demonstrate that the orbital states of the spinel oxide $FeV_2O_4$, which possesses active orbital degrees of freedom on both Fe and V ions, are uniquely resolved by combining valence electron density (VED) analysis based on state-of-the-art synchrotron x-ray diffraction with spin-polarized density-functional-theory calculations. Our results reveal that temperature-dependent rearrangements of orbital occupations drive successive structural transitions that accompany collinear and noncoplanar ferrimagnetic ordering, establishing a direct correspondence between orbital anisotropy and spin structure. This work shows that experimentally determined VED provides a decisive real-space constraint on competing theoretical solutions, offering a powerful and broadly applicable framework for elucidating the microscopic mechanisms of complex phase transitions in strongly correlated electron systems.

## I. INTRODUCTION

The anisotropy of valence electrons is fundamentally intertwined with a variety of emergent quantum phenomena, including charge and orbital order [1–5], unconventional superconductivity [6,7], quantum spin liquid [8], and multipole order [9]. In such systems, orbital configuration directly governs the electronic states and physical properties. However, since external fields that couple directly to orbital degrees of freedom are limited, orbital states are considerably less accessible to direct experimental observation than lattice distortions or magnetic structures. Traditionally, orbital states have been inferred indirectly from Jahn–Teller distortions, which manifest as structural deformations of metal–ligand polyhedra through electron-phonon coupling. While such structural information generally provides valuable insight in strongly correlated electron systems, it is rarely sufficient on its own. Electron correlation, relativistic spin–orbit interaction, and finite-temperature effects act on comparable energy scales, obscuring the simple correspondence between lattice distortions and orbital occupations. Consequently, the orbital state cannot be unambiguously determined solely from crystallographic analysis, necessitating direct experimental probes of valence electron anisotropy.

A prototypical example in which orbital degrees of freedom give rise to exotic electronic states is the spinel-type iron vanadium oxide $FeV_2O_4$, which crystallizes in the space group $Fd\bar{3}m$ at high temperatures [Fig. 1(a)]. In this compound, each $Fe^{2+}$ ion ($3d^6$) is coordinated by four $O^{2-}$ ions to form a $FeO_4$ regular tetrahedron, while each $V^{3+}$ ion ($3d^2$) is surrounded by six $O^{2-}$ ions to form a $VO_6$ octahedron with trigonal distortions. In contrast to other spinel vanadates $AV_2O_4$ ($A$ = Mg, Zn, Cd, Mn) [10–15], not only V but also Fe ions host orbital degeneracies in the high-temperature cubic phase in $FeV_2O_4$.

Figure 1(b) shows the changes in lattice deformations across the successive phase transitions in $FeV_2O_4$ [15–18]. Hereafter, $a$, $b$, and $c$ denote the lattice parameters in the standard crystallographic setting, whereas $a'$, $b'$, and $c'$ are used as a unified notation, referenced to the cubic phase, for all temperature-dependent phases. As shown in Fig. 1(b), upon cooling, the cubic lattice first transforms to tetragonal with $I4_1/amd$ symmetry at $T_s$ = 140 K, in which the lattice is contracted along the $c'$-axis. Below $T_{N1}$ = 110 K, collinear ferrimagnetic (C-FM) order emerges, with $Fe^{2+}$ and $V^{3+}$ moments aligned antiparallel to each other [Fig. 1(c)], accompanied by a further lowering of lattice symmetry from tetragonal to orthorhombic, with space group $Fddd$. The ferrimagnetic moment appears along the longest $a'$-axis [Fig. 1(b)].

Below $T_{N2}$ = 70 K, the structure reverts from orthorhombic to tetragonal with space group $I4_1/amd$, accompanied by a transition to a noncollinear ferrimagnetic (NC-FM) state [15–18]. In contrast to the high-temperature tetragonal phase, the low-temperature tetragonal lattice is elongated along the fourfold axis, unlike in other spinel vanadates $AV_2O_4$ with orbitally inactive $A$ ions [10–15]. Note that Fig. 1(b) depicts the low-temperature tetragonal phase using a face-centered tetragonal cell with $a'$ as the unique axis. A previous neutron diffraction study suggests that the Fe spins remain collinear, whereas the V spin moments are tilted by approximately 55° from the direction antiparallel to the Fe spin at the lowest temperature [Fig. 1(c)] [17]. In the two lower-temperature phases, the spin arrangements of Fe and V are strongly coupled to their orbital states [15,16,18]. Hereafter, the four phases from high to low temperature are referred to as Cubic, HT-tetra, Ortho, and LT-tetra.

Despite extensive experimental [15–25] and theoretical [26–32] investigations of $FeV_2O_4$, the microscopic origins of the successive phase transitions remain under debate. Because strong on-site Coulomb interactions open an energy gap on the order of electron volts, the low-energy configurations of spin and orbital are affected by Kugel-Khomskii-type interactions between adjacent metal sites [33], relativistic spin–orbit interaction, and Jahn–Teller interaction. The competition among them introduces theoretical ambiguity, giving rise to several possible thermal equilibrium states depending on the parameters [26,27,29,34].

We previously visualized the valence electron density (VED) distribution in the Cubic phase of $FeV_2O_4$ [35] by combining single-crystal synchrotron x-ray diffraction with the core differential Fourier synthesis (CDFS) method [36,37]. This approach enabled us to experimentally determine the Fe and V 3*d* wavefunctions (details are provided below) and to clarify the orbitally degenerate states in the high-temperature Cubic phase, which had long remained controversial from a theoretical standpoint. A central issue in $FeV_2O_4$, however, is how these orbital states evolve through the successive structural and magnetic transitions upon cooling. This question has remained unresolved, mainly because symmetry lowering induces multi-domain states that hinder precise data acquisition.

In this study, we address this problem by applying an external magnetic field to a smaller crystal across the magnetic phase transitions. In the low-temperature phases, this approach strongly biases the domain populations, reducing the number of populated domains to two or fewer and thereby allowing us to obtain high-quality diffraction data suitable for precise VED analysis. This enables us to track the temperature evolution of the VED across successive phase transitions and to establish a comprehensive experimental picture of the associated reconfiguration of Fe and V orbital states. Comparison between the experimental VED and spin-polarized density-functional-theory (DFT) calculations further allows us to distinguish between nearly degenerate microscopic solutions for the ground-state orbital configuration and to clarify their relation to the spin structure [Figs. 1(d) and 1(e)]. More broadly, our results show that experimentally determined VED provides a practical route to identifying orbital states and resolving their coupling to spin and lattice degrees of freedom in strongly correlated electron systems.

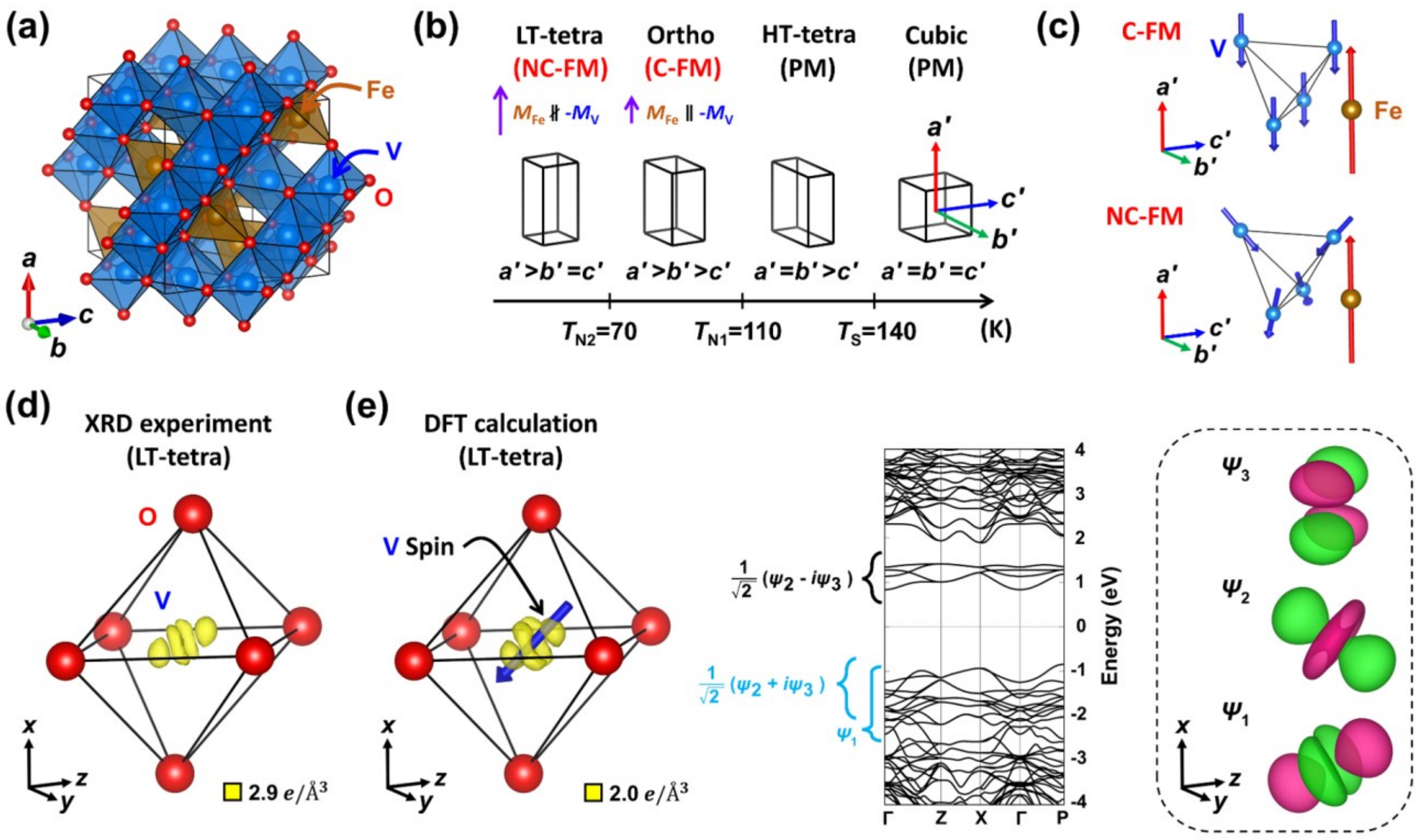

Fig.1. Structural and magnetic phase transitions in $FeV_2O_4$, and comparison between x-ray diffraction experiments and DFT calculations. (a) Crystal structure of $FeV_2O_4$ in the Cubic phase, consisting of $FeO_4$ tetrahedra and $VO_6$ octahedra. (b) Change in lattice deformations across the successive phase transitions. Purple arrows indicate the direction of magnetization ($M$). The magnetic states are denoted as paramagnetic (PM), collinear ferrimagnetic (C-FM), and noncollinear ferrimagnetic (NC-FM). (c) Schematic illustration of the magnetic structures in the Ortho (C-FM) and LT-tetra (NC-FM) phases. The Fe and V spins are shown by thick red and blue arrows, respectively. (d) Experimental VED distribution around a $VO_6$ octahedron at 50 K in the LT-tetra phase. The yellow color represents the iso-density surface of the VED. (e) VED distribution and electronic band dispersion in the LT-tetra phase, computed using spin-polarized DFT+$U$ calculations with spin–orbit coupling. The dashed box shows three wavefunctions ($\Psi_1$, $\Psi_2$, and $\Psi_3$) originating from the $t_{2g}$ orbitals, where green and pink represent the positive and negative signs of the wavefunctions, respectively. The VED distribution, corresponding to the occupied $t_{2g}$ orbitals below the Fermi level, is shown around the V sites, exhibiting anisotropy similar to that observed experimentally. The blue arrow represents the spin on the V site. The $\boldsymbol{x}$, $\boldsymbol{y}$, and $\boldsymbol{z}$ axes in (d) and (e) represent the local coordinate systems defined by $\boldsymbol{x} \| \boldsymbol{a'}$, $\boldsymbol{y} \| \boldsymbol{b'}$, and $\boldsymbol{z} \| \boldsymbol{c'}$ for Fe at (1/8, 1/8, 1/8) and V at (1/2, 1/2, 1/2), respectively (see Supplemental Material, Section 1 [38]).

## II. METHODS

### A. X-ray diffraction

Single-crystal x-ray diffraction experiments were conducted at the BL02B1 beamline [40] of the SPring-8 synchrotron facility in Japan. The single crystals used in this study were the same as those in Ref. [15]. Temperature was controlled using an $N_2$/He gas-blowing device. The wavelengths were 0.30945 Å at 160 and 50 K, and 0.31080 Å at 130 and 100 K. Prior to the measurements in the Ortho and LT-tetra phases, the sample was cooled while applying a magnetic field along the *a*'-axis of the cubic phase using a Neodymium magnet to align structural domains. The magnet was removed before the X-ray diffraction measurements, and the diffraction data were collected without an external magnetic field. Further details are provided in the Supplemental Material, Section 1 [38]. Diffraction patterns were recorded using a two-dimensional CdTe PILATUS detector [41] with a dynamic range of ~$10^6$. Data were acquired using the Fine Slice method by oscillating the crystal and dividing reciprocal space into intervals of $\Delta\omega$ = 0.1° [42]. The intensities of Bragg reflections were collected by CrysAlisPro [43]. Intensities of equivalent reflections were averaged using SORTAV [44], and structural parameters were refined by Jana2006 [45]. Here, by utilizing only high-angle reflections ($\sin\theta/\lambda > 0.6$ Å$^{-1}$), where the contribution of spatially spread valence electrons to x-ray diffraction is negligible, structural parameters including atomic displacement parameters were obtained with high accuracy.

### B. CDFS analysis

The CDFS method was used to extract the VED distribution. The [Ar]-type configuration ($1s^2$,$2s^2$,$2p^6$,$3s^2$,$3p^6$) of the Fe and V atoms as well as $1s^2$ electrons of O atoms were regarded as core electrons. The contribution of the thermal vibration was subtracted from the VED using the atomic displacement parameters determined by the high-angle analysis. The voxel size of the three-dimensional VED distribution was 0.05 Å$^3$. We used the STO-COPPENS Slater-type orbital (STO) library implemented in Jana2006 for the CDFS analysis. The crystal structure and VED distributions were visualized using VESTA [46].

### C. DFT calculation

The DFT electronic structure calculations for $FeV_2O_4$ were performed using Quantum Espresso [47] using the experimental structure of the LT-tetra phase obtained in this study. We employed relativistic norm-conserving pseudopotentials with Perdew-Burke-Ernzerhof (PBE) [48] exchange-correlation functional, which were taken from the PseudoDojo [49]. We performed spin-polarized calculations considering the spin–orbit coupling with $\boldsymbol{k}$-mesh of 7×7×7. The effects of Coulomb interactions of Fe 3*d* and V 3*d* orbitals were incorporated within the DFT+$U$ formalism [50] with the Hubbard $U$ parameters of 5 eV. The energy cutoff was set to 100 Ry for the wave functions and 400 Ry for the charge density. Based on the DFT electronic structure, we constructed Wannier orbitals [51,52] using RESPACK [53,54] and Wannier90 [55] for Fe 3*d*, V 3*d*, and O 2*p* manifold. Since a primitive unit cell consists of two Fe sites, four V sites, and eight O sites, the number of Wannier orbitals was (5×2+5×4+3×8)×2=108, where the factor of two accounted for the summation over spin degrees of freedom.

## III. RESULTS and DISCUSSION

### A. VED in the Cubic phase

Figure 2(a) shows the VED distribution at 160 K (Cubic phase), obtained from the CDFS analysis. As shown by yellow iso-density surfaces of 3.9$e$/Å$^3$, the VED distribution appears isotropic around the $O^{2-}$ ions, which is consistent with the $2s^2 2p^6$ electronic configuration. On the other hand, weak and pronounced anisotropies are observed around the Fe and V sites, respectively, as shown in the zoomed-in views [Figs. 2(b) and 2(c)]. Since the $FeO_4$ forms a regular tetrahedron, the Fe $3d$ orbitals split into a lower-lying $e$ doublet and a higher-lying $t_2$ triplet. In the high-spin configuration for $Fe^{2+}$ ($3d^6$), the five majority-spin electrons fully occupy the $3d$ orbitals, while the remaining minority-spin electron resides in degenerate $e$ orbitals ($3z^2–r^2$ and $x^2–y^2$) with equal probability [inset, dashed square in Fig. 2(b)]. Since x-ray diffraction probes orbitals as time- and space-averaged states, the observed VED anisotropy around the Fe site can be interpreted as a manifestation of this orbital disorder [35].

A similar interpretation can be applied to the VED distribution around the $V^{3+}$ ($3d^2$) ion. In a regular octahedron, the $3d$ orbitals would be split into a higher-lying $e_g$ doublet and a lower-lying $t_{2g}$ triplet. However, the $VO_6$ octahedron is slightly elongated along the local threefold axis, reducing the local symmetry of the V site to $.\bar{3}m$. As a result, the $t_{2g}$ orbitals are further split into $a_{1g}$ and $e_g^{'}$ orbitals. Considering only the crystal field from the adjacent six $O^{2-}$ ions forming the octahedron, the $e'_g$ orbitals would lie lower in energy than the $a_{1g}$ orbital. The Ewald method [56], however, predicts that the influence of more distant ions reverses this hierarchy, stabilizing the $a_{1g}$ orbital relative to the $e'_g$ orbitals [35]. The observed VED anisotropy around the V site confirms the latter situation, where one electron occupies the $a_{1g}$ orbital and the other is equally distributed to the $e'_g$ doublet [Fig. 2(c)]. Starting from this Cubic phase orbital state, we discuss how successive phase transitions lift the degeneracy of Fe and V orbitals and lead to the ground state.

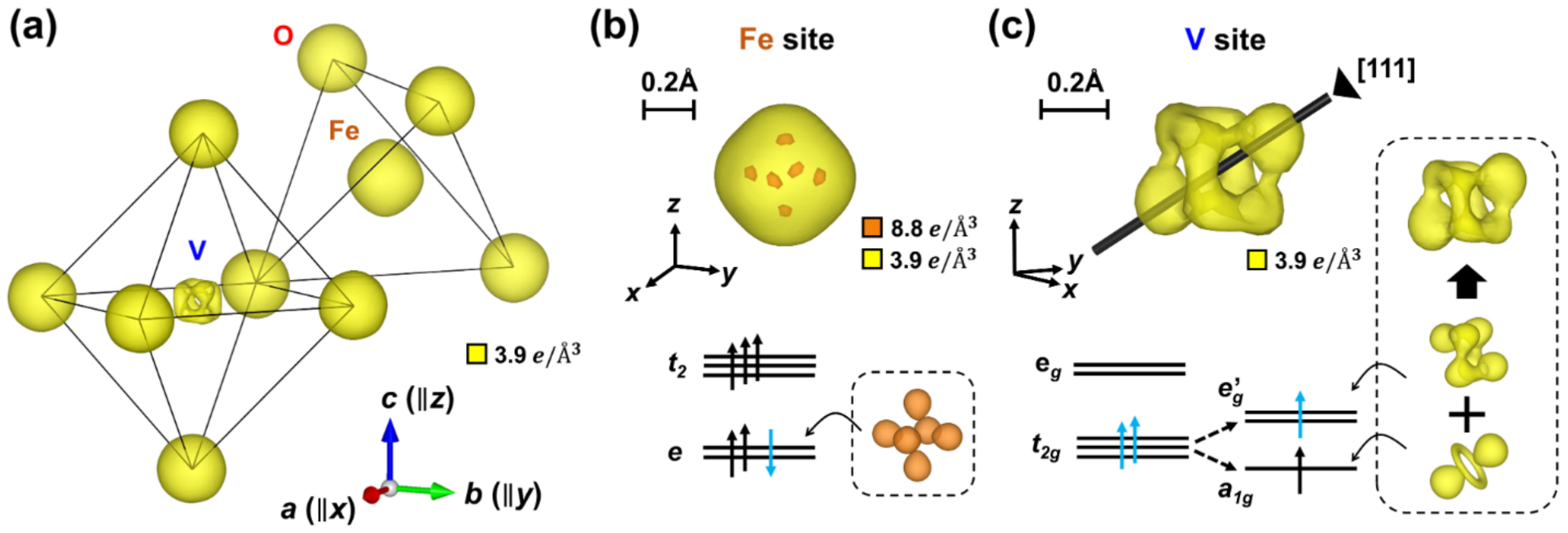

Fig.2. VED distribution of $FeV_2O_4$ in the Cubic phase. (a) Experimental VED distribution at 160 K, showing a single $FeO_4$ tetrahedron and a single $VO_6$ octahedron sharing an O atom. Iso-density surfaces at a level of $3.9e/Å^3$ are shown by yellow. (b) Enlarged VED distribution around the Fe site and schematic of $Fe^{2+}$ $3d^6$ high-spin configuration. VED distribution of a down-spin in the doubly degenerate $e$ orbitals, shown in orange, is calculated as the average of electron densities of the $3z^2–r^2$ and $x^2–y^2$ orbitals. (c) Enlarged VED distribution around the V site and schematic of $V^{3+}$ $3d^2$ configuration. A black stick with a black triangle represents the three-fold axis. The trigonal distortion splits the $t_{2g}$ orbital into a lower energy $a_{1g}$ and higher energy doubly degenerate $e_g^{'}$ orbitals. VED distribution is calculated as the sum of electron densities of $a_{1g}$ and the average of $e_g^{'}$ orbitals. Spins with an orbital degree of freedom are depicted as light blue arrows, while other spins are shown by black arrows.

## B. $FeO_4$ deformation and VED around the Fe site

Figure 3 shows the distortions of $FeO_4$ and $VO_6$ polyhedra in each phase, as determined by synchrotron x-ray diffraction. As shown in Fig. 3(a), the $FeO_4$ distortion contains two-dimensional normal modes of $Q_{2,Fe}$ and $Q_{3,Fe}$, which cause splitting of the $e$ orbitals [Fig. 2(b)]. First, we examine the successive change in $FeO_4$ distortion. Figures 3(b) and 3(c) present the temperature dependence of $Q_{2,Fe}$ and $Q_{3,Fe}$. The orange iso-density surfaces in Fig. 3(c) show the minority-spin electron densities of Fe predicted from the local $FeO_4$ distortions in the strong electron-lattice coupling limit. In the Cubic phase, the $Q_{2,Fe}$ and $Q_{3,Fe}$ modes are absent. In the HT-tetra phase at 130 K, negative $Q_{3,Fe}$ coupled to the $3z^2–r^2$ orbital emerges, followed by the emergence of $Q_{2,Fe}$ in the Ortho phase at 100 K. Here, we consider the structural domain with $Q_{2,Fe} > 0$. In the LT-tetra phase, the $FeO_4$ tetrahedron is elongated along the $x$-axis and stabilizes the $y^2–z^2$ orbital.

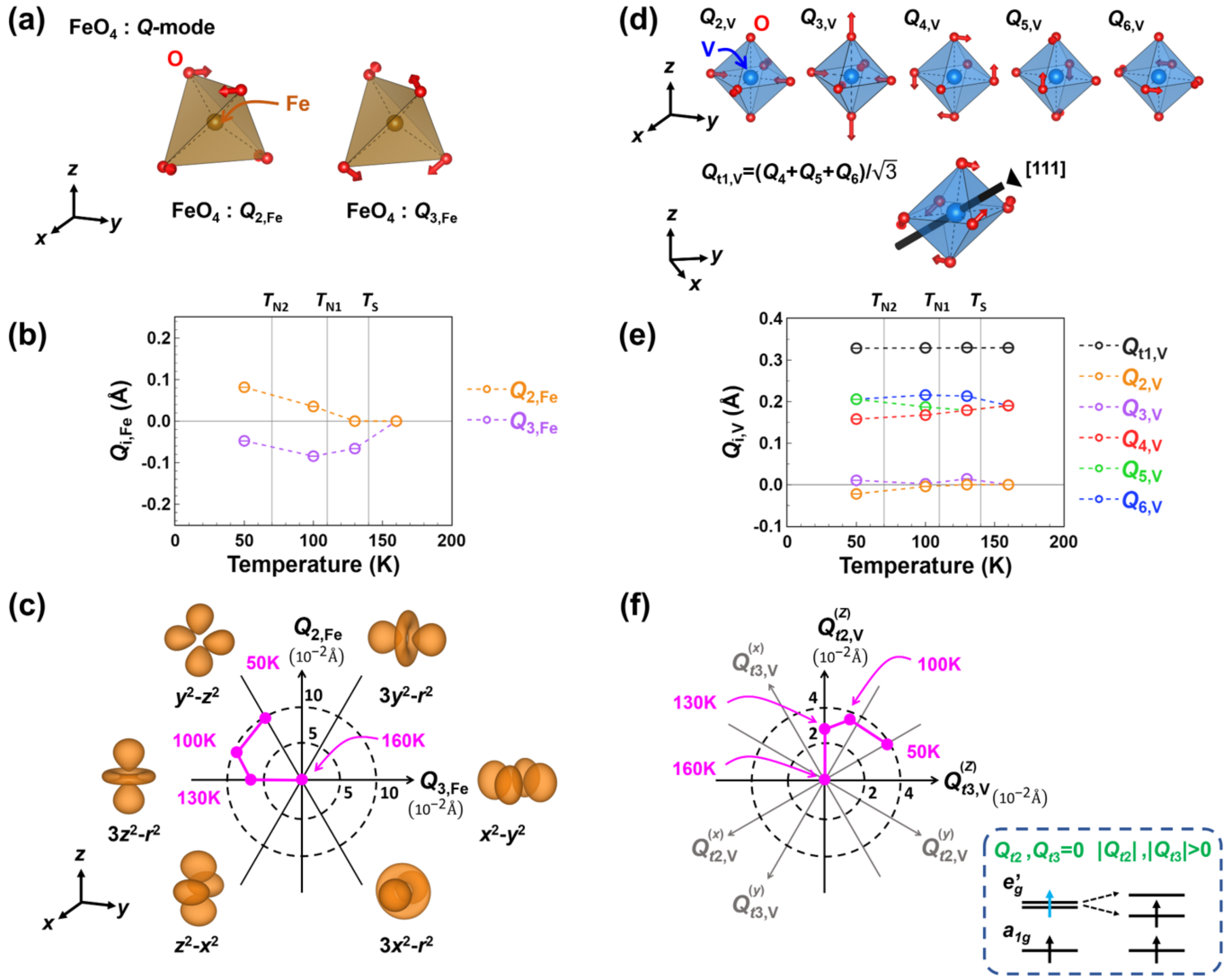

Fig.3. Distortions of $FeO_4$ and $VO_6$ associated with the phase transitions. (a) Schematic illustration of the normal mode in the $FeO_4$ tetrahedron. The distortion can be represented using the $Q_{2,\mathrm{Fe}}$ and $Q_{3,\mathrm{Fe}}$ modes. In the $FeO_4$ tetrahedron, $Q_{4,\mathrm{Fe}} = Q_{5,\mathrm{Fe}} = Q_{6,\mathrm{Fe}} = 0$. (b) Temperature dependence of $Q_{2,\mathrm{Fe}}$ and $Q_{3,\mathrm{Fe}}$. (c) Temperature dependence of $Q_{2,\mathrm{Fe}}$ and $Q_{3,\mathrm{Fe}}$ on a two-dimensional plane. This diagram also illustrates which 3*d* orbital is stabilized through coupling with the distortion of $FeO_4$. (d) Schematic illustration of *Q*-modes of $VO_6$ distortion. In the case of octahedral coordination, symmetric distortion is decomposed into six modes: $A_{1g}$ ($Q_{1,\mathrm{V}}$: breathing), $E_g$ ($Q_{2,\mathrm{V}}$, $Q_{3,\mathrm{V}}$), $T_{2g}$ ($Q_{4,\mathrm{V}}$, $Q_{5,\mathrm{V}}$, $Q_{6,\mathrm{V}}$) modes. The trigonal distortion is represented as $Q_{t1,\mathrm{V}} = (Q_{4,\mathrm{V}} + Q_{5,\mathrm{V}} + Q_{6,\mathrm{V}})/\sqrt{3}$. (e) Temperature dependence of the $Q_{i,\mathrm{V}}$ ($i$=2 to 6) and $Q_{t1,\mathrm{V}}$ values. (f) Temperature dependence of $Q^{(z)}_{t2,\mathrm{V}} = (-Q_{4,\mathrm{V}} - Q_{5,\mathrm{V}} + 2Q_{6,\mathrm{V}})/\sqrt{6}$ and $Q^{(z)}_{t3,\mathrm{V}} = (-Q_{4,\mathrm{V}} + Q_{5,\mathrm{V}})/\sqrt{2}$ modes, which are orthogonal to $Q_{t1,\mathrm{V}}$, on a two-dimensional plane.

To directly elucidate Fe 3*d* orbital states, we examine the VED distribution. Figure 4(a) shows the temperature evolution of the VED around the Fe site. In the Cubic phase, high-density regions, shown by orange, are located on the *x*-, *y*-, and *z*-axes. In the HT-tetra phase, enhanced VED develops along the *z*-axis reflecting the splitting of the *e* orbitals and minority-spin occupation of $3z^2–r^2$ orbital. Upon further cooling into the Ortho phase, the VED along the *y*-axis increases, described by a linear combination of the $3z^2–r^2$ and $y^2–z^2$ orbitals. In the LT-tetra phase, high-density regions extending within the $yz$-plane are observed, corresponding to the stabilization of the $y^2$–$z^2$ orbital. Figure 4(b) presents the calculated VED distribution for the minority-spin of Fe, which reproduces the anisotropy observed in the high iso-density surfaces shown in orange [Fig. 4(a)]. These results are consistent with those expected from the distortion of the $FeO_4$ tetrahedron [Fig. 3(c)].

The successive changes in Fe 3$d$ orbital state can be understood in terms of competition between the cooperative Jahn–Teller effect below $T_S$ and the spin–orbit coupling in the magnetic phases, which stabilize the $3z^2$–$r^2$ and $y^2$–$z^2$ orbitals, respectively [15,16,18,57]. The effective Hamiltonians for the Jahn–Teller coupling and the spin–orbit coupling in the $FeO_4$ tetrahedron are given by $H_{JT}=A(\tau_x Q_{2,Fe}-\tau_z Q_{3,Fe})$, and $H_{SOC} = B/6[(3S_z^2 - S^2)\tau_z - \sqrt{3}(S_x^2 - S_y^2)\tau_x]$ [58], respectively. Here, $A$ (<0) is an electron-phonon coupling constant, $\tau_z$ and $\tau_x$ are pseudospin operators representing $3z^2$–$r^2$ ($\tau_z = 1/2$) and $x^2$–$y^2$ ($\tau_z = -1/2$) orbitals. $S$, $S_x$, $S_y$, and $S_z$ are spin operators, and $B$ (>0) represents a second-order perturbation of spin–orbit coupling. Firstly, we examine the HT-tetra phase. The anharmonic term $U = A_3Q^3\cos(3\theta)$ in the elastic energy favors the compressed tetragonal distortion. Here, $A_3$ (>0) is the third-order elastic constant, $Q = (Q_{2,Fe}{}^2 + Q_{3,Fe}{}^2)^{1/2}$, and $\theta$ is the polar angle of the distortion vector ($Q_{3,Fe}$, $Q_{2,Fe}$) in the $Q_{2,Fe}$- $Q_{3,Fe}$ plane. The resulting cooperative Jahn–Teller distortion stabilizes the Fe $3z^2$-$r^2$ orbitals. At lower temperatures where magnetic order emerges, the system undergoes a further structural transition. Under second-order spin-orbit coupling, the favored orbital depends on the spin orientation: the $y^2$–$z^2$ orbital is stabilized when the spin moment is aligned along the $x$-axis. When $Fe^{2+}$ magnetic order emerges at $T_{N1}$, the orbital state is therefore modulated by spin–orbit coupling. As the spins align along the $x$-axis below $T_{N1}$ in the Ortho phase, the $y^2$–$z^2$ orbital is stabilized by spin–orbit coupling, leading to a hybrid of $3z^2$–$r^2$ and $y^2$–$z^2$ orbitals [15]. Consequently, $Q_{2,Fe}$ becomes non-zero through Jahn–Teller coupling, resulting in an orthorhombic distortion. Here, $Q_{2,Fe} > 0$, as we adopt the orthorhombicity of $a > b > c$. Indeed, in $FeCr_2O_4$, where $Cr^{3+}$ ions without orbital degrees of freedom occupy the octahedral sites, the ground state remains an orthorhombic phase with conical order of Fe and Cr spins [58,59]. This suggests that the cooperative Jahn–Teller effect and spin–orbit coupling at the Fe site account for the tetragonal-to-orthorhombic transition. However, the reentrant tetragonal symmetry in the ground state of $FeV_2O_4$ cannot be explained within this Fe-only framework, indicating an essential role of the orbital degrees of freedom at the V sites.

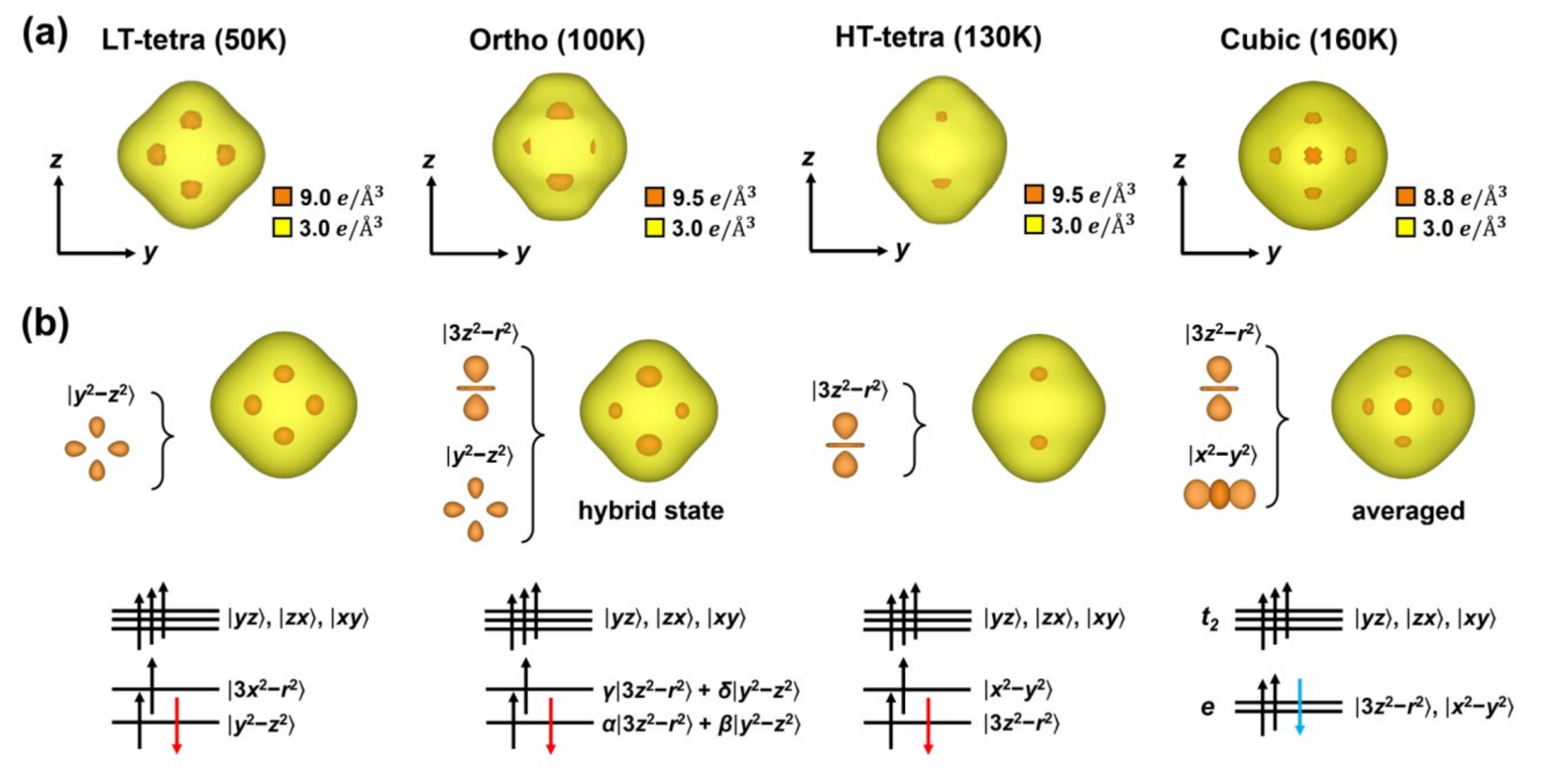

Fig.4. VED distribution around the Fe site and its orbital configuration. (a) Temperature dependence of the observed VED distributions around the Fe site. (b) VED distributions reconstructed from the fitted Fe 3*d* wave functions and schematic diagrams of the high-spin $3d^6$ configuration. A minority-spin electron retaining orbital degeneracy is depicted by a light-blue arrow, minority-spin electrons whose degeneracy is lifted in the low-temperature phases are depicted by red arrows, and majority-spin electrons are depicted by black arrows. In the Ortho phase, the two basis states of the *e* orbitals are expressed as linear combinations of $|3z^2\text{-}r^2\rangle$ and $|y^2\text{-}z^2\rangle$, $\alpha|3z^2\text{-}r^2\rangle + \beta|y^2\text{-}z^2\rangle$ and $\gamma|3z^2\text{-}r^2\rangle + \delta|y^2\text{-}z^2\rangle$. Here, the coefficients $\alpha, \beta, \gamma, \delta$ are obtained from a basis transformation of the orthonormal states $c_1|3z^2\text{-}r^2\rangle + c_2|x^2\text{-}y^2\rangle$ and $-c_2|3z^2\text{-}r^2\rangle + c_1|x^2\text{-}y^2\rangle$ with $|c_1|^2+|c_2|^2=1$. Using relation $|y^2\text{-}z^2\rangle = -\frac{\sqrt{3}}{2}|3z^2\text{-}r^2\rangle - \frac{1}{2}|x^2\text{-}y^2\rangle$, the coefficients are given by $\alpha = c_1 - \sqrt{3}c_2$, $\beta = -2c_2$, $\gamma = -\sqrt{3}c_1 - c_2$, and $\delta = -2c_1$. The best fit to the experimental observation is obtained with $\alpha = 1.97$, $\beta = 1.89$, $\gamma = 0.37$, and $\delta = -0.66$. As $|3z^2\text{-}r^2\rangle$ and $|y^2\text{-}z^2\rangle$ are not mutually orthogonal, the transformed coefficients $\alpha$, $\beta$, $\gamma$, and $\delta$ do not satisfy a simple squared-sum normalization; the normalization condition is $\alpha^2 + \beta^2 - \sqrt{3}\alpha\beta = 1$ for $\alpha|3z^2\text{-}r^2\rangle + \beta|y^2\text{-}z^2\rangle$, and similarly for $\gamma$ and $\delta$.

## C. $VO_6$ deformation and VED around the V site

The symmetric distortion of a $VO_6$ octahedron can be decomposed into $E_g$ ($Q_{2,\mathrm{V}}$, $Q_{3,\mathrm{V}}$) and $T_{2g}$ ($Q_{4,\mathrm{V}}$, $Q_{5,\mathrm{V}}$, $Q_{6,\mathrm{V}}$) modes, corresponding to variations in the V-O bond lengths and the O-V-O bond angles, respectively [Fig. 3(d)]. In the Cubic phase, $Q_{4,\mathrm{V}} = Q_{5,\mathrm{V}} = Q_{6,\mathrm{V}} > 0$ and their linear combination $Q_{t1,\mathrm{V}} = (Q_{4,\mathrm{V}} + Q_{5,\mathrm{V}} + Q_{6,\mathrm{V}})/\sqrt{3}$ corresponds to elongation or compression of the $VO_6$ octahedron along the [111] axis. This trigonal distortion leads to the splitting of the $t_{2g}$ into $a_{1g}$ and $e'_g$ orbitals. Figure 3(e) shows the temperature dependence of each $Q$ mode of the $VO_6$ octahedron. While the $Q_{2,\mathrm{V}}$ and $Q_{3,\mathrm{V}}$ values, which directly couple to the splitting of the $e_g$ orbitals, remain nearly zero, $Q_{4,\mathrm{V}}$, $Q_{5,\mathrm{V}}$, and $Q_{6,\mathrm{V}}$ values exhibit pronounced temperature dependence, reflecting their coupling to the splitting of the $t_{2g}$ orbitals. Notably, the large $Q_{t1,\mathrm{V}}$ value is nearly temperature-independent, indicating that one electron continues to occupy the $a_{1g}$ orbital down to the lowest temperatures. To further probe the splitting of the $e'_g$ orbitals, we examine the temperature dependence of $Q^{(z)}_{t2,V} = (-Q_{4,V} - Q_{5,V} + 2Q_{6,V})/\sqrt{6}$ and $Q^{(z)}_{t3,V} = (-Q_{4,V} + Q_{5,V})/\sqrt{2}$ modes, which are orthogonal to $Q_{t1,V}$ [Fig. 3(f)]. In the Cubic phase, both $Q^{(z)}_{t2,V}$ and $Q^{(z)}_{t3,V}$ values are exactly zero, indicating that the $e'_g$ orbitals remain degenerate. In the HT-tetra phase at 130 K, a positive $Q^{(z)}_{t2,V}$ mode emerges while $Q^{(z)}_{t3,V}$ remains zero, corresponding to tetragonal distortion. Subsequently, in the Ortho phase at 100 K, a $Q^{(z)}_{t3,V}$ mode appears. Note that $Q^{(z)}_{t3,V}$ and $Q^{(z)}_{t2,V}$ have the same sign. Upon entering the LT-tetra phase, $Q^{(z)}_{t3,V}$ gradually increases. We emphasize that the value at 50 K lies along the $Q^{(x)}_{t2,V} = (-Q_{5,V} - Q_{6,V} + 2Q_{4,V})/\sqrt{6}$ axis when the $Q_{4,V}$ direction is taken as the principal axis, corresponding to $Q^{(x)}_{t3,V} = (-Q_{5,V} + Q_{6,V})/\sqrt{2} = 0$. The continuous evolution from positive $Q^{(z)}_{t2,V}$ to negative $Q^{(x)}_{t2,V}$ should reflect modifications of the 3$d$ wave functions. Although the amplitudes of $Q_{t2,V}$ and $Q_{t3,V}$ are small, these modes provide structural signatures of the symmetry lowering that lifts the $e_g'$ degeneracy. Their small amplitudes also indicate that the V orbital state is not determined by the $VO_6$ distortion alone but is stabilized cooperatively by the Fe–V exchange interaction, spin–orbit coupling, on-site Coulomb interaction, and ligand field of distorted lattice. Therefore, we next examine the VED distribution around the V site to identify the orbital state realized in this spin–orbital–lattice–coupled environment.

Next, we turn to the VED distribution around the V site [Fig. 5(a)], which also continuously changes upon cooling. Here, we define $t_{2g}$ wavefunctions as $\Psi_1 = C_1|yz\rangle + C_2|zx\rangle + C_3|xy\rangle$, $\Psi_2 = D_1|yz\rangle + D_2|zx\rangle + D_3|xy\rangle$, and $\Psi_3 = E_1|yz\rangle + E_2|zx\rangle + E_3|xy\rangle$, with $|C_1|^2+|C_2|^2+|C_3|^2=1$, $|D_1|^2+|D_2|^2+|D_3|^2=1$, $|E_1|^2+|E_2|^2+|E_3|^2=1$, where the coefficients are chosen to satisfy the site-symmetry and mutual-orthogonality constraints (see Supplemental Material, Section 3 [38]). To elucidate the orbital states, the quantum parameters $C_i$, $D_i$, and $E_i$ ($i$ = 1 to 3) are optimized by fitting the calculated VED distributions to reproduce the observed anisotropy. Assuming that the trigonal distortion is large enough for one electron to occupy $\Psi_1$, the calculated $3d^2$ VED is expressed as

$$\rho_{\mathrm{calc}}(\boldsymbol{r}) = |\psi_1|^2 + \eta|\psi_2|^2 + (1-\eta)|\psi_3|^2 \tag{1}$$

where $\eta$ represents the electron filling of $\Psi_2$. The evaluation function $s$ for the fitting is defined as

$$s = \frac{\sum_r |\rho_{\mathrm{obs}}(\boldsymbol{r}) - \kappa^3 \rho_{\mathrm{calc}}(\kappa \boldsymbol{r})|}{\sum_r |\rho_{\mathrm{obs}}(\boldsymbol{r})|} \tag{2}$$

where $\rho_{\mathrm{obs}}(\boldsymbol{r})$ is the observed VED around the V site. In general, hybridization between neighboring atoms may modify the radial distribution of the observed VED from that of an isolated-atom model. To account for this effect, a radial scaling parameter $\kappa$ is introduced. The quantum parameters are optimized by minimizing the evaluation function in Eq. (2) within a radial range of $0.25 < r < 0.6$ Å around the V site (see Supplemental Material, Fig. S9 [38]). The evaluation function is minimized at approximately $\eta = 0.5$ in all the phases. Figure 5(b) shows the calculated VED distributions with the optimized quantum parameters. The temperature dependence of the observed VED anisotropy is attributed to the reconfiguration of the $\Psi_1$ wavefunction [Fig. 5(c)], since $\Psi_2$ and $\Psi_3$ are determined by the $C_i$ parameters through the site-symmetry constraints and the orthogonality conditions with respect to $\Psi_1$ (see Supplemental Material, Section 3 [38]). The observed temperature evolution of the V orbital state is closely coupled to the distortion of the $VO_6$ octahedra. The large $Q_{t1,V}$-mode distortion corresponding to the trigonal elongation of $VO_6$ inherent in the spinel structure suggests its dominant effect on the $t_{2g}$ orbital state, which is neglected in previously proposed theoretical models [26,27,29,34].

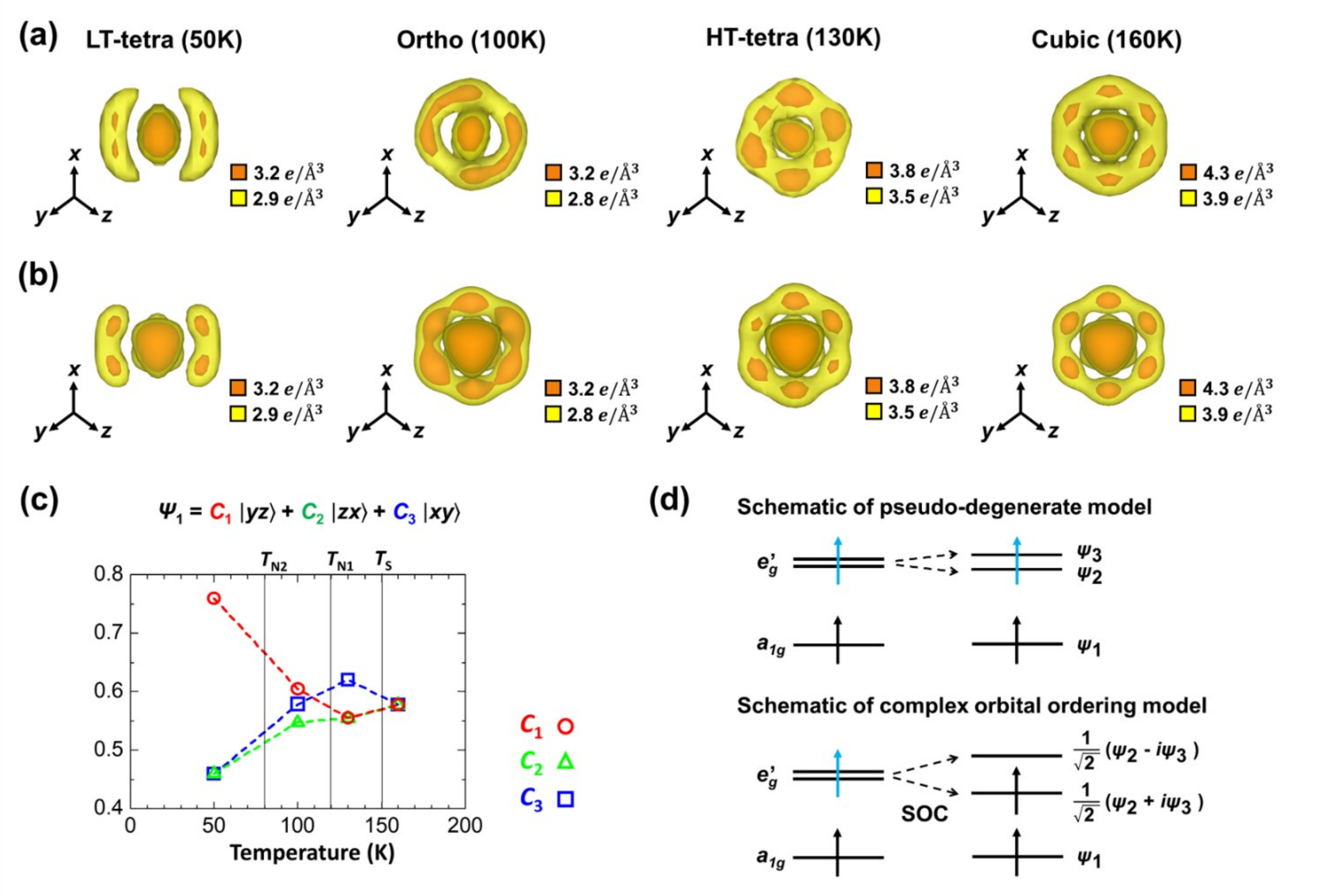

Fig.5. Comparison between observed and calculated VED distributions around the V site. (a) Temperature dependence of the observed VED distributions around the V site, viewed along the [111] axis. (b) VED distribution reconstructed from the fitted V 3$d$ wave functions. The parameters were obtained by fitting to the experimental VED distribution. (c) Temperature dependence of quantum parameters $C_1$, $C_2$, and $C_3$ of the $\Psi_1$ wavefunction. (d) Two possible models for the 3$d^2$ orbital state of the $V^{3+}$ ion.

## D. Spin-polarized DFT calculations in the LT-tetra phase

The 3$d^2$ orbital states of $V^{3+}$ in the low-temperature phases with $\eta \sim 0.5$ can be interpreted within two possible frameworks, as shown in Fig. 5(d). One possible scenario is that the splitting of the $e_g'$ orbitals is sufficiently small that the system resides in a thermally accessible pseudo-degenerate state. Alternatively, the VED with $\eta = 0.5$ can also arise from wavefunctions stabilized in magnetically ordered states through relativistic spin–orbit interaction [27]. The two scenarios are impossible to distinguish based solely on the VED distribution.

To address this issue, we perform spin-polarized DFT+$U$ calculations for the LT-tetra phase considering the relativistic spin–orbit interaction. The three V $t_{2g}$ wavefunctions are defined as $\Psi_1=\gamma\,|yz\rangle+\sqrt{(1-\gamma^2)/2}|zx\rangle+\sqrt{(1-\gamma^2)/2}|xy\rangle$, $\Psi_2=-\sqrt{1-\gamma^2}|yz\rangle+(\gamma/\sqrt{2})|zx\rangle+(\gamma/\sqrt{2})|xy\rangle$, and $\Psi_3=(1/\sqrt{2})|zx\rangle-(1/\sqrt{2})|xy\rangle$ [as illustrated in Figs. 1(e) and 6(c)]. The Hubbard $U$ raises the energy of unoccupied $e_g$ states, thereby promoting gap formation between the $t_{2g}$ and $e_g$ orbitals. Interestingly, we find two solutions depending on the mechanism of orbital order. Note that, because these two solutions are nearly degenerate in energy, a unique determination cannot be made based on DFT calculations alone.

In one solution, Fe spin moments are ferromagnetically arranged along the four-fold axis ($-x$ direction) and V spins are aligned nearly in the opposite direction with small canting arising from the spin–orbit interaction [Fig. 6(a)]. The Hubbard $U$ effectively enhances the crystal field effect of the compression of $VO_6$ along the $x$-axis, raising the energy of the $\Psi_3$ orbital, composed of $zx$ and $xy$ orbitals, relative to the $\Psi_2$ orbital, which contains the $yz$ component [Figs. 6(b) and 6(c)]. Consequently, orbital ordering consisting of $\Psi_1$ and $\Psi_2$ emerges. The predicted VED is shown in Fig. 6(d), which is, however, inconsistent with the experimental observation [Fig. 5(a)].

In the other solution, the V spin moments exhibit a “two-in-two-out” configuration within the $V_4$ tetrahedra of the pyrochlore network [Fig. 6(e)]. In this case, $\Psi_1$ and $(\Psi_2+i\Psi_3)/\sqrt{2}$ constitute the orbital order [Figs. 6(f) and 6(g)]. The complex orbital $(\Psi_2+i\Psi_3)/\sqrt{2}$ gives an orbital angular moment of approximately 1 $\mu_B$, oriented close to $\langle 111\rangle$, corresponding to the direction toward the center of the $V_4$ tetrahedron. This orientation is nearly opposite to the local spin directions in the two-in-two-out configuration. In fact, an x-ray magnetic circular dichroism study [24] has demonstrated that the V site possesses a sizable orbital magnetic moment component parallel to the magnetization. It is likely that spin–orbit interaction triggers orbital ordering, assisted by the gap enhancement due to the Hubbard $U$. In this solution, spin–orbit coupling selects the complex orbital state within the $e_g'$ doublet, whereas the Hubbard $U$ enhances the separation between the occupied and unoccupied V 3$d$ states. The local SOC energy scale can be estimated from an effective spin–orbital coupling within the $e'_g$ doublet. For the related spinel vanadate $MnV_2O_4$, a recent microscopic estimate [60] gives $|\lambda_{\|}|=22.7$ meV for the on-site spin–orbital coupling $\lambda \parallel \tau^Z S^Z$ in the effective $e'_g$ doublet, corresponding to approximately 260 K. Although this value was obtained for $MnV_2O_4$, it provides a useful reference for the expected energy scale in the $V^{3+}$ $t_{2g}$ orbitals of $FeV_2O_4$. This energy scale is significantly larger than the thermal energy at $T_{N2}\approx 70$ K, $k_BT\approx 6$ meV, suggesting that spin–orbit coupling can play an important role in selecting the complex orbital state. Figure 6(h) displays the calculated VED, which shows good agreement with the observed VED [Fig. 5(a)].

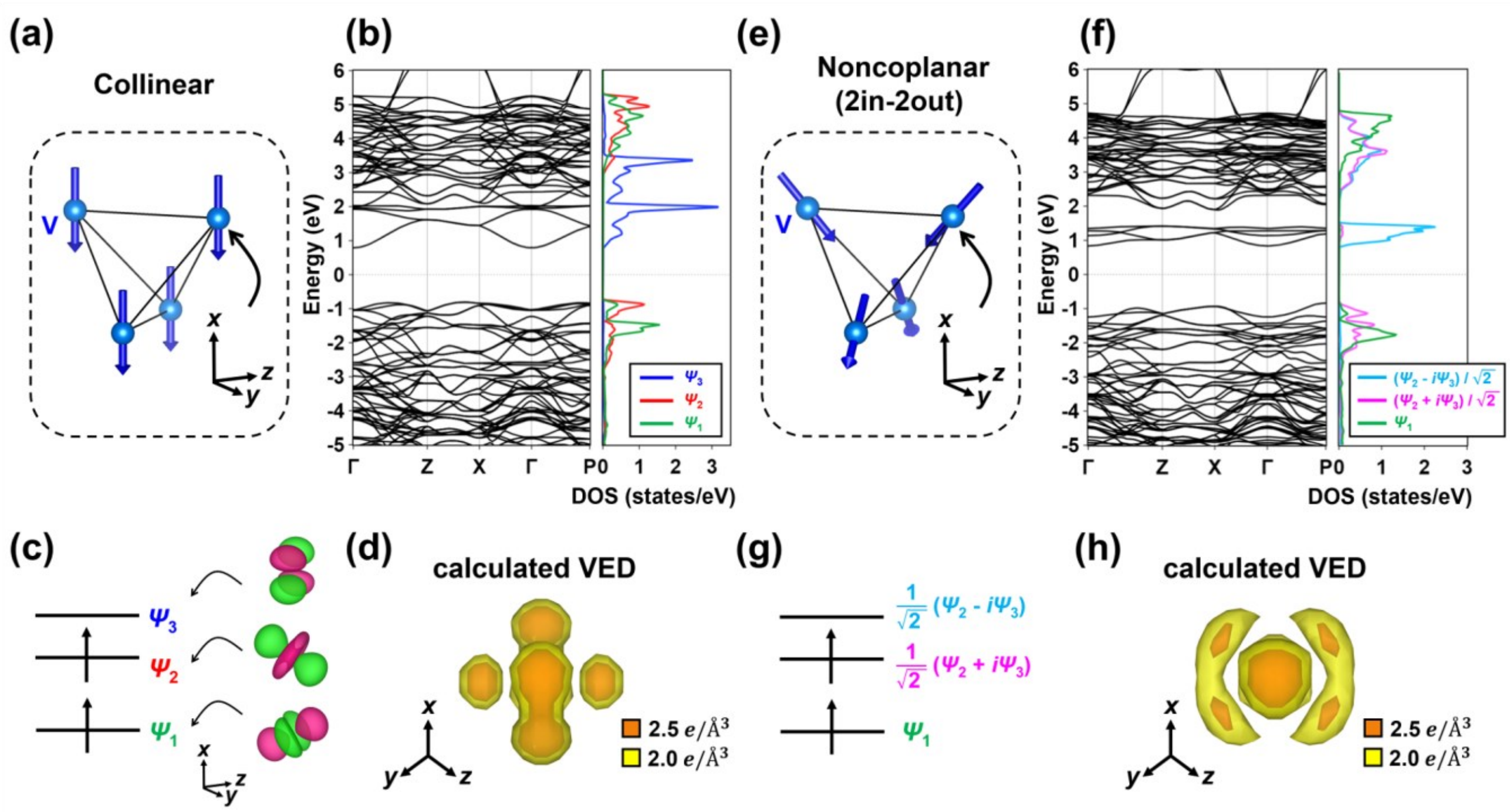

Fig.6. Results for spin-polarized DFT+$U$ calculations with spin-orbit coupling in the LT-tetra phase. (a) Magnetic structure in the collinear solution, where the V spins are antiferromagnetically aligned with the Fe spins. (b) Band structure along the $\boldsymbol{k}$-path of $\Gamma(0,0,0)\rightarrow Z\left(\frac{2\pi}{a},0,0\right)\rightarrow X\left(\frac{\pi}{a},\frac{\pi}{a},0\right)\rightarrow\Gamma(0,0,0)\rightarrow P\left(\frac{\pi}{a},\frac{\pi}{a},\frac{\pi}{c}\right)$ and projected density of states of the V $t_{2g}$ orbitals for the collinear solution. (c), (d) Schematic diagram of the orbital states, corresponding to the collinear solution and calculated VED distribution at the V site. (e) Magnetic structure in the noncoplanar solution with two-in-two-out type V magnetic moments. In this solution, the spin of V is canted approximately 43 deg. from the -$x$ direction. (f) Band structure and projected density of states of the V $t_{2g}$ orbitals for the noncoplanar solution. (g), (h) Schematic diagram of the orbital states, corresponding to the noncoplanar solution and calculated VED distribution at the V site. Here, the projected density of states of the V $t_{2g}$ orbitals was obtained from the band structure using the wave function $\Psi_1=\gamma\,|yz\rangle+\sqrt{(1-\gamma^2)/2}|zx\rangle+\sqrt{(1-\gamma^2)/2}|xy\rangle$, $\Psi_2=-\sqrt{1-\gamma^2}|yz\rangle+\left(\gamma/\sqrt{2}\right)|zx\rangle+(\gamma/\sqrt{2})|xy\rangle$, and $\Psi_3=(1/\sqrt{2})|zx\rangle-(1/\sqrt{2})|xy\rangle$. The parameter $\gamma$=0.76, determined by fitting to the experimental results, was used for both the collinear and noncoplanar solutions.

## E. Expected noncoplanar arrangement of V spins

Importantly, in the wavefunction $\Psi_1=\gamma|yz\rangle+\sqrt{(1-\gamma^2)/2}|zx\rangle+\sqrt{(1-\gamma^2)/2}|xy\rangle$, where $\gamma=0.76$ is determined by fitting to the experimental VED, the occupation of the $yz$ orbital is inequivalent to those of the $zx$ and $xy$ orbitals. The latter two carry equal weights, as dictated by the calculated symmetry. As a result, the threefold rotational symmetry is explicitly broken, whereas the exchange symmetry between $zx$ and $xy$ remains preserved. As a consequence, the local crystal field retains a mirror symmetry that leaves the $yz$ orbital invariant and exchanges $zx$ and $xy$. This mirror plane contains the local ⟨111⟩ axis, and therefore the orbital angular momentum induced by spin–orbit coupling is constrained to lie within this plane. The spin moment is thus constrained within the same mirror plane, fixing its azimuthal direction. Using the orbital wavefunctions in the LT-tetra phase, the

direction of the local orbital moment is obtained as $(\gamma,\ \sqrt{(1-\gamma^2)/2},\ \sqrt{(1-\gamma^2)/2})$. This corresponds to a polar angle of approximately 41 degrees from the $+x$ direction and about 14 degrees from the direction parallel to the local $\langle 111 \rangle$ axis. This estimate of the local orbital-moment direction roughly agrees with the V-spin canting angle of 43 degrees obtained by the DFT+$U$+SOC calculation, which may infer a non-negligible role of the relativistic spin–orbit coupling at V site in the LT-tetra phase. Since the canting direction is confined to the local mirror plane, the spin configuration on the pyrochlore lattice becomes intrinsically noncoplanar, thereby generating scalar spin chirality. More generally, carriers endowed with an angular momentum may experience an effective magnetic field arising from the scalar spin chirality, which may give rise to unique transport phenomena such as the magnon Hall effect. These results suggest intertwined spin and orbital degrees of freedom. The observed orbital-ordering pattern provides an important clue to the spin configuration of the system and, more importantly, a key to identifying the underlying mechanism of the orbital ordering. The role of the V orbital state discussed here is therefore to constrain the local magnetic anisotropy through spin–orbit coupling. In the following section, we discuss how this V spin–orbital coupling contributes to the recovery of tetragonal lattice symmetry via the $VO_6$ distortion.

### F. Possible origin of the reentrant tetragonal ground state

Although we cannot yet provide a complete microscopic explanation for the reentrant tetragonal ground state, the experimentally established facts allow us to propose the following physically consistent scenario for the sequence of phase transitions. Throughout the following discussion, we distinguish between the global crystallographic axes ($a'$, $b'$, and $c'$), which describe the macroscopic lattice distortion, and the local quantization axes at the Fe and V sites, which determine the orbital and spin states through the local crystal field and spin–orbit coupling. Structural distortions are therefore described using the global crystallographic axes, whereas orbital occupations and spin–orbit coupling are discussed in terms of the local electronic coordinate systems.

**(i) Cubic → HT-tetra transition at $T_S$**

The structural transition from the Cubic phase to the HT-tetra phase at $T_S$ is governed primarily by the Fe orbital–lattice coupling. At $T_s$, the $Fe^{2+}$ $e$-orbital degeneracy is lifted by the cooperative Jahn–Teller effect. As discussed in Sec. III C, when an anharmonic term in the elastic energy is considered, the compressed tetragonal distortion of the $FeO_4$ tetrahedron is more stable than the elongated tetragonal distortion, and the $3z^2-r^2$ orbital is stabilized.

**(ii) HT-tetra → Ortho transition at $T_{N1}$**

Upon cooling below $T_{N1}$, ferrimagnetic order further modifies the Fe orbital state through the second-order perturbation of the Fe-site spin–orbit coupling. When the Fe spin is aligned along the local $x$ direction, the $y^2-z^2$ orbital becomes energetically favorable. Consequently, the Fe orbital state evolves from the pure $3z^2-r^2$ configuration to a mixed state of $3z^2-r^2$ and $y^2-z^2$, producing the orthorhombic distortion. The local $x$ axis of the $FeO_4$ tetrahedron approximately corresponds to the crystallographic $a'$ axis. Therefore, stabilization of the $y^2-z^2$ orbital leads to elongation of the $FeO_4$ tetrahedron along the global $a'$ axis, giving rise to the orthorhombic lattice with $a' > b' > c'$.

This interpretation is fully consistent with the VED analysis. The Fe-site electron-density anisotropy changes markedly across both $T_S$ and $T_{N1}$, whereas the V-site anisotropy changes only weakly. These results demonstrate that the structural evolution from the Cubic phase through the HT-tetra phase to the Ortho phase is governed primarily by the Fe orbital–lattice coupling.

**(iii) Ortho → LT-tetra reentrant transition at $T_{N2}$**

The reentrant structural transition from the Ortho phase to the LT-tetra phase at $T_{N2}$ has a fundamentally different origin. Unlike the higher-temperature transitions, which are driven by the Fe orbital–lattice coupling, this transition is triggered by the local spin–orbital state of the V ions.

Below $T_{N2}$, the V-site spin–orbit coupling gains in importance. The quenched orbital angular momentum partially recovers to couple to the spin moment, preferring higher symmetry of ligand field. This suggests that the LT-tetra phase provides a more favorable environment for the spin–orbital coupled state than the lower-symmetry Ortho phase. In the LT tetra phase, the V-site has $..2/m$ local symmetry, which is absent in the Ortho phase. In fact, the DFT calculation suggests that the V spin moment lies in the local mirror plane.

The V-site spin–orbit coupling affects the balance between the anharmonicity in $FeO_4$ distortion and Fe-site spin–orbit coupling, which is the origin of the orthorhombic lattice distortion, and results in the recovery of the tetragonal lattice. Notably, the distortion amplitude of the $FeO_4$ tetrahedra is substantially larger than that of the $VO_6$ octahedra [Figs. 3(c) and 3(f)], indicating that the $FeO_4$ units dominate the magnitude of the macroscopic lattice deformation. Therefore, although the reentrant transition is initiated by the local V spin–orbital state, the accompanying structural distortion is amplified primarily through the Fe orbital–lattice instability.

In summary, the sequence of structural transitions is governed by two distinct mechanisms. The Cubic → HT-tetra and HT-tetra → Ortho transitions are driven primarily by the Fe orbital–lattice coupling, whereas the Ortho → LT-tetra reentrant transition is triggered by the V spin–orbital coupling. The cooperative interaction between these two mechanisms stabilizes the reentrant tetragonal ground state elongated along the global $a'$ axis.

Finally, we consider the orbital states of the $V^{3+}$ ion in the intermediate HT-tetra and Ortho phases. Previous thermal conductivity measurements [22] and powder neutron diffraction [17] suggest that V orbital order is not established in these intermediate phases. Consequently, the electron is likely to fluctuate within the Bloch sphere defined by $\Psi_2$ and $\Psi_3$ orbitals. If such fluctuations are sufficiently slow on the relevant experimental time scale, they may give rise to a dynamical Jahn–Teller effect. However, we observe neither x-ray diffuse scattering nor anomalies in anisotropic atomic displacement parameters (see Supplemental Material, Section 1 [38]). To directly probe this pseudo-degenerate state in the intermediate phases, spectroscopic investigations of phonon dynamics, such as Raman or infrared spectroscopy, would therefore be useful.

## IV. CONCLUSION

We have directly visualized the temperature evolution of the VED distributions in $FeV_2O_4$ across its successive structural and magnetic phase transitions. The observed VED anisotropies around the Fe and V sites reveal how the local orbital states are reconstructed from the orbitally degenerate cubic phase to the low-temperature phases, demonstrating that the phase evolution is governed by cooperative spin–orbital–lattice coupling involving both sublattices. By combining these real-space observations with spin-polarized DFT calculations, we further demonstrate that the lowest-temperature V orbital state can be uniquely identified through direct comparison with experiment, thereby discriminating between competing microscopic scenarios that are nearly degenerate in theory. These results establish experimentally determined VED as a decisive probe of orbital physics in strongly correlated materials, going beyond conventional crystal-structure-based inference. More broadly, our work provides a general strategy for identifying hidden orbital states and their coupling to magnetism and lattice symmetry in quantum materials, particularly when multiple metastable solutions compete on comparable energy scales.

**Acknowledgements**

We thank A. Nakano, T. Ohashi, Y. Yamanaka, and T. Sasaki for supporting the x-ray diffraction experiments, and T. Katsufuji, T. Hasegawa, K. Siratori, and T. Nakai for the fruitful discussions. This work was supported by JST SPRING (Grant No. JPMJSP2108), JSPS KAKENHI (Grant No. 22K14010, 23H04869, 24H01644, 25H01246 and 25H01252), RIKEN TRIP initiative (Advanced General Intelligence for Science Program, Many-body Electron Systems), JST-ASPIRE (Grant No. JPMJAP2512), and JST FOREST (Grant No. JPMJFR2362). The synchrotron radiation experiments were performed at SPring-8 with the approval of the Japan Synchrotron Radiation Research Institute (JASRI) (Proposal No.2022A1246, 2022A1247, 2023A1520, 2023B2006, 2024B0304, 2025A1505 and 2025A1732).

**Data availability**

The data that support the findings of this study are available from the corresponding author upon reasonable request.